\documentstyle[12pt,twoside]{article}

\def\lsim{\mathrel{\rlap{\lower4pt\hbox{\hskip1pt$\sim$}}
    \raise1pt\hbox{$<$}}}         
\def\gsim{\mathrel{\rlap{\lower4pt\hbox{\hskip1pt$\sim$}}
    \raise1pt\hbox{$>$}}}         
\newlength{\dinwidth}            
\newlength{\dinmargin}           
  \newenvironment{defl}[1]%
  {\begin{list}{}{\settowidth{\labelwidth}{#1}%
  \setlength{\leftmargin}{\labelwidth}%
  \addtolength{\leftmargin}{\labelsep}%
  \setlength{\itemsep}{0pt plus 1pt}
  \setlength{\parsep}{0pt plus 1pt}
  \setlength{\topsep}{0pt plus 1pt}
  \setlength{\partopsep}{0pt plus 1pt}
  \setlength{\parskip}{1mm plus 1mm minus 1mm}
  }}%
  {\end{list}}
\begin{document}
\thispagestyle{empty}
\noindent
DESY 96--059        \hfill ISSN 0418-9833\\
April 1996  
\begin{center}
  \begin{Large}
  \begin{bf}
MAJOR 1.5 -- A Monte Carlo Generator for \\
Heavy Majorana Neutrinos in $ep$ Collisions\\
  \end{bf}
  \end{Large}
  \vspace{5mm}
  \begin{large}
J. Rathsman$^a$, G. Ingelman$^{a,b}$\\
  \end{large}
  \vspace{3mm}
rathsman@tsl.uu.se ~~~~ ingelman@desy.de\\
  \vspace{3mm}
$^a$ Dept. of Radiation Sciences, Uppsala University,
Box 535, S-751 21 Uppsala, Sweden\\
$^b$ Deutsches Elektronen-Synchrotron DESY,
Notkestrasse 85, D-22603 Hamburg, FRG\\
  \vspace{5mm}
\end{center}
\begin{quotation}
\noindent
{\bf Abstract:}
The Monte Carlo generator {\sc Major} 1.5 simulates the production and
decay of heavy Majorana neutrinos via lepton mixing or exchange of
`light'  right-handed $W$-bosons in deep inelastic scattering, i.e.
$e^{\pm} p \rightarrow {N} X \rightarrow{e}^{\pm} W^{\mp} X$ or 
$\nu_e Z X$. Physics and programming aspects are described in 
this manual.
\end{quotation}
%
%
\section{Physics of included processes}
The standard model of electroweak and strong interactions has been
remarkably successful in describing the experimental data. Nevertheless,
it cannot be the ultimate theory due to its theoretical shortcomings,
{\it e.g.} for the understanding of the mass and family structure of quarks
and leptons and the many free parameters. 
Attempts to solve the theoretical problems of the standard model
have been made based on various theoretical grounds and several
extended theories have been suggested and studied in detail.
These theories are normally based on some larger symmetry group
which unifies the interactions and is spontaneously broken down to
the standard model gauge group.
Particular attention has been given to models with an additional
$U(1)$ symmetry or with left-right symmetry in the form
$SU(2)_L\otimes SU(2)_R\otimes U(1)_{B-L}$,
which can both be subgroups of the unification groups $SO(10)$ and
$E_6$. The questions of neutrino masses and lepton number
violation, which are of general importance (for a review see \cite{review}),
enter explicitly in this
context and, in particular, heavy Majorana neutrinos may be present
\cite{BG}.

\begin{figure}[tbh]
 \vspace{5.5cm}
 \includegraphics{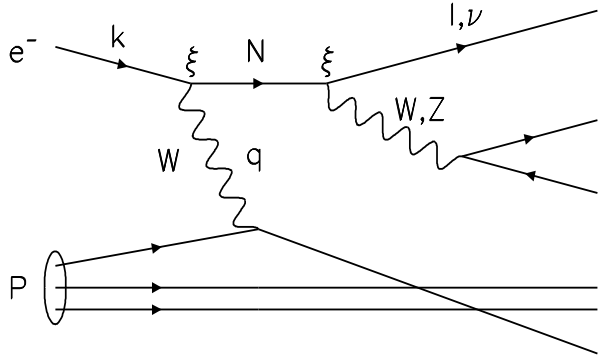}
 \caption[]{{\it
 Production (in $ep$ collisions) and decays of heavy Majorana neutrinos.
 }}
 \label{feyn}
\end{figure}

The simplest process, in terms of minimal new physics assumptions, to produce
heavy Majorana neutrinos in $ep$ collisions is through normal charged current 
interactions as illustrated in Fig.~\ref{feyn}. 
The cross-section is suppressed, not only by limited phase space due to the
large neutrino mass $m_N$, but also by the required mixing $\xi$ 
between light and heavy Majorana neutrinos \cite{BG,proc}. 
The light Majorana neutrinos can be identified with the normal neutrinos
$\nu_e , \nu_\mu$ and $\nu_\tau$, whose small masses is naturally explained
through the see-saw mechanism. 
The mixing in the leptonic 
part of the charged current interaction is given by a unitary 
Kobayashi-Maskawa type matrix $V$ and the matrix $\xi$. 
A power series in $\xi$ gives the connection between the weak eigenstates 
$\nu_L, \nu_R$ and the Majorana mass eigenstates $\nu , N$ \cite{BG}, 
{\it i.e.} 
\begin{equation}
\nu_L=\frac{1-\gamma_5}{2}(\nu+\xi{N}+...)\: ,\: \: \:
\nu_R=\frac{1+\gamma_5}{2}(N-\xi^T\nu+...). 
\nonumber
\end{equation}

The heavy Majorana neutrinos will decay into either 
$\nu_{\ell}Z^0$, ${\ell}^+W^-$ or ${\ell}^-W^+$ where the weak bosons can be
taken as being on-shell since we only consider Majorana neutrino masses 
above the boson masses ($m_N>m_B$) due to mass limits \cite{majorep}.
Using the narrow width approximation for the heavy Majorana neutrino propagator
in Fig.~\ref{feyn}, the cross-section for the process 
$e^-p\to NX$ where $N\to \ell^\pm W^\mp$ or $N\to \nu_{\ell} Z^0$
is \cite{BG} 
\begin{eqnarray} \label{dsigma}
\frac{d\sigma}{dx\,dy\,dp_{\ell\perp}\,dy_{\ell}} & = &
\frac{G_F^2m_W^4|(V\xi)_{eN}|^2}{32\sqrt{2}\pi^3\hat{s}m_N\Gamma_N}
\frac{m_W^2}{(y\hat{s}+m_W^2)^2}
\frac{1}{m_{N\perp}} \nonumber \\
 & & \times\left[\cosh(y_N-y_{\ell})-
\frac{m_N^2-m_B^2-2p_{\ell\perp}p_{N\perp}}
{2m_{N\perp}p_{\ell\perp}}\right]^{-1/2} \nonumber \\
 & & \times\left[\frac{m_N^2-m_B^2+2p_{\ell\perp}p_{N\perp}}
{2m_{N\perp}p_{\ell\perp}}
-\cosh(y_N-y_{\ell})\right]^{-1/2} \nonumber \\
 & & \times\left\{A\cdot\left[u(x,\mu^2)+c(x,\mu^2)\right] +
B\cdot\left[\overline{d}(x,\mu^2)+\overline{s}(x,\mu^2)\right]\right\}
\end{eqnarray}
For incoming positron the last row in Eq.(\ref{dsigma}) is replaced by
\begin{eqnarray} 
\times\left\{C\cdot\left[d(x,\mu^2)+s(x,\mu^2)\right] +
D\cdot\left[\overline{u}(x,\mu^2)+\overline{c}(x,\mu^2)\right]\right\}
\end{eqnarray}

Another possibility to produce right-handed neutrinos is through the exchange
of right-handed $W$-bosons, $W_R$, see e.g. \cite{BGR}. 
If it is assumed that the $W_R$ is 
lighter than the heavy neutrino and that there is no mixing with standard model
gauge bosons then the same cross-section formula can be used
except that $G_F^2m_W^4|(V\xi)_{eN}|^2=g_L^2|(V\xi)_{eN}|^2$ is 
replaced by the right handed gauge coupling, $g_R^2$, and everywhere else 
$m_W$ is replaced by $m_{W_R}$. 
To simplify the picture one may also assume
that the heavy neutrino only decays according to the charged decay 
modes, $N\rightarrow \ell^\pm W_R^\mp$,
thus avoiding any extra assumptions on the $Z^{\prime}$-mass.
Finally, if only two-body decays are considered, then $W_R$ only decays 
into quark-antiquark pairs due to the larger mass of the right-handed neutrino.
It is difficult to obtain model-independent limits on $m_{W_R}$ and
according to \cite{Langacker} the $W_R$ can be as light as the ordinary
$W_L$ if extreme fine-tuning is allowed.

The independent variables are chosen as the normal deep inelastic 
scaling variables $x$ and $y$ together with the transverse
momentum  $p_{\ell\perp}$ and rapidity $y_{\ell}$ of the final state lepton
$\ell$ from the Majorana neutrino decay. With four-momenta as in Fig.~1, 
$s = (P + k)^{2}$, $\hat{s}=xs$, $W^2=(P+q)^2$ and $Q^{2} = - q^{2}$.
As usual, $x=Q^{2}/2 P \cdot q$ and $y=P \cdot q/P \cdot k$, 
but they are globally constrained to $\frac{m_{N}^{2}}{s} \le x \le 1$ and
$0 \le y \le 1- \frac{m_{N}^{2}}{\hat{s}}$.
The phase space is given by
\begin{eqnarray}
\label{xlim}
\frac{1}{s} \, \left( p_{N \perp}^{min} +
m_{N \perp}^{min} \right)^{2} \leq &x& \leq 1 \\
\label{ylim}
\frac{\hat{s} - m_{N}^{2} - \sqrt{(\hat{s} - m_{N}^{2})^{2} - 4 \,
\hat{s} \, p_{N \perp}^{min \, 2}}}{2 \, \hat{s}}
\leq &y& \leq
\frac{\hat{s} - m_{N}^{2} + \sqrt{(\hat{s} - m_{N}^{2})^{2} - 4 \,
\hat{s} \, p_{N \perp}^{min \, 2}}}{2 \, \hat{s}} \,\\
\label{ptlim}
0 \leq & p_{\ell\perp} & \leq \frac{m_N^2-m_B^2}{2m_N^2}\sqrt{s}
\end{eqnarray}
\vspace{-0.6cm}
\begin{eqnarray}
\label{rlim}
\frac{m_N^2-m_B^2-2p_{\ell\perp}p_{N\perp}}{2m_{N\perp}p_{\ell\perp}} \leq & 
\cosh(y_N-y_{\ell}) & \leq
\frac{m_N^2-m_B^2+2p_{\ell\perp}p_{N\perp}}{2m_{N\perp}p_{\ell\perp}}
\end{eqnarray}

The indices $N$ and $W$ denote the Majorana neutrino and the 
$W$ boson exchanged in the $t$-channel, whereas $B$ and $\ell$ denote 
the weak boson and lepton from the Majorana neutrino decay, respectively.
The symbols $y_{\ell}$, $y_N$, $m$, $m_\bot =\sqrt{m^2+p_\bot^2}$ and $p_\bot$ 
refer to rapidity, mass, transverse mass and transverse momentum, respectively.
All frame dependent quantities are understood to be in the lab frame.
$\Gamma_N$ is the total width of the Majorana neutrino and 
the functions $u$, $c$, $d$ and $s$ are the parton density
functions in the proton evaluated at a suitable scale $\mu$.   

The coefficient functions $A,B$ depend on the decay products of the 
Majorana neutrino and are for the charged decays given by \cite{BG}
\begin{eqnarray} 
\label{A-+}
A({\ell^-}W^+) & = & |(V\xi)_{{\ell}N}|^2\frac{4\hat{s}}{m_W^2}
\nonumber \\ & & \left[(m_N^2-m_W^2)(2m_W^2\hat{s}-m_N^4)
-2m_N^2(2m_W^2-m_N^2)p_{\ell\perp}(xE_pe^{-y_{\ell }}+E_ee^{y_{\ell}})\right]\\
\label{B-+}
B({\ell^-}W^+) & = & |(V\xi)_{{\ell}N}|^2\frac{8
(\hat{s}(1-y)-m_N^2)}{m_W^2} \nonumber \\ & &
\left[\hat{s}(1-y)m_W^2(m_N^2-m_W^2)-
m_N^2(2m_W^2-m_N^2)p_{\ell\perp}xE_pe^{-y_{\ell }}\right] \\
\label{A+-}
A({\ell^+}W^-) & = & |(V\xi)_{{\ell}N}|^2\frac{4\hat{s}m_N^2}{m_W^2}
\nonumber \\ & & \left[(m_N^2-m_W^2)(\hat{s}-2m_W^2)
+ 2(2m_W^2-m_N^2)p_{\ell\perp}(xE_pe^{-y_{\ell }}+E_ee^{y_{\ell}})\right] \\
\label{B+-}
B({\ell^+}W^-) & = & |(V\xi)_{{\ell}N}|^2
\frac{4m_N^2(\hat{s}(1-y)-m_N^2)}{m_W^2}
\nonumber \\ & & \left[\hat{s}(1-y)(m_N^2-m_W^2)+2(2m_W^2-m_N^2)p_{\ell\perp}
xE_pe^{-y_{\ell }}\right]
\end{eqnarray}
and for the neutral decay $N\rightarrow\nu_{\ell}Z$ by \cite{majorep} 
\begin{eqnarray}
\label{A00}
A({\nu_{\ell}}{Z}) & = & |\xi_{\nu_{\ell}N}|^2
\frac{2\hat{s}(\hat{s}-m_N^2)(m_N^2-m_Z^2)(m_N^2+2m_Z^2)}
{\cos^2\theta_Wm_Z^2} \\
\label{B00}
B({\nu_{\ell}}{Z}) & = & |\xi_{\nu_{\ell}N}|^2
\frac{2\hat{s}(1-y)(\hat{s}(1-y)-m_N^2)
(m_N^2-m_Z^2)(m_N^2+2m_Z^2)}
{\cos^2\theta_Wm_Z^2}
\end{eqnarray}
For incoming positron the coefficient functions $C,D$ are related to $A,B$ in
the following way,
\begin{eqnarray} 
D({\ell^+}W^-) & = & A({\ell^-}W^+) \\
C({\ell^+}W^-) & = & B({\ell^-}W^+) \\
D({\ell^-}W^+) & = & A({\ell^+}W^-) \\
C({\ell^-}W^+) & = & B({\ell^+}W^-) \\
D({\nu_{\ell}}{Z}) & = & A({\nu_{\ell}}{Z})\\
C({\nu_{\ell}}{Z}) & = & B({\nu_{\ell}}{Z})
\end{eqnarray}
The width of the heavy Majorana neutrino is given by
\begin{eqnarray} \label{Gamma-tot}
\Gamma_N & = & 
\sum_{\ell} \left[
2\Gamma(N \to \ell^{\pm} W^{\mp}) + \Gamma(N \to \nu_{\ell} Z) 
\right] \nonumber \\ & = & 
\sum_{\ell}\frac{G_F}{8\sqrt{2}\pi m_N^3}
\left[2|(V\xi)_{{\ell}N}|^2(m_N^2-m_W^2)^2(m_N^2+2m_W^2)\right. 
\nonumber \\ & &
\hspace{2.5cm} \: + \: 
\left. |\xi_{\nu_{\ell}N}|^2(m_N^2-m_Z^2)^2(m_N^2+2m_Z^2)\right].
\end{eqnarray}
\begin{figure}[htb]
 \vspace{6.5cm}
 \includegraphics{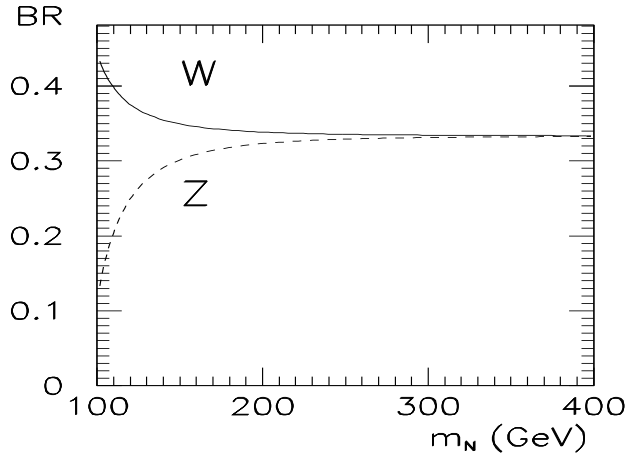}
 \caption[]{{\it
 Branching ratios for $N\to \ell^{\pm}W^{\mp}$ (full curve)
 and  $N\to \nu_{\ell}Z^0$ (dashed curve) as a function of the Majorana
 neutrino mass $m_N$.
 }}
 \label{branch}
\end{figure}
Assuming the Kobayashi-Maskawa type matrix $V$ for the lepton sector to be
diagonal, so that $|(V\xi)_{{\ell}N}|=|\xi_{\nu_{\ell}N}|$, 
we obtain the branching ratios shown in Fig.~\ref{branch}. 
Although all three are 
1/3 at large Majorana neutrino masses, for masses not far above the $W$ and
$Z$ masses there is a substantial phase space suppression of the decay into 
the $Z$.

For a detailed study on present limits on Majorana neutrino masses and mixings,
as well as their phenomenology and suitable search strategies at $ep$ colliders 
we refer to \cite{majorep}.
\section{The Monte Carlo implementation}
The importance sampling method is used to generate phase space
points according to the complete differential cross-section formula
given by Eq.~(\ref{dsigma}). 
In short, the importance sampling method can be described as a variable 
transformation,
\begin{equation}
(x, y, p_{\ell\perp}, y_{\ell})\: \rightarrow 
\:\left( H_x(x),H_y(y),
H_{p_{\ell\perp}}({p_{\ell\perp}}),H_{y_{\ell}}({y_{\ell}}) \right)
\end{equation}
where
$H_x(x)=\int^x\!{h}_{x}(x^\prime)dx^\prime$, etc. 
and $h$ are the so called weighting functions.
A phase-space point, $(x, y, p_{\ell\perp}, y_{\ell})$ is first chosen from 
the inverse of the integrals, {\it e.g.} 
\begin{equation}
x=H_x^{-1}[H_{x,\min}+R(H_{x,\max}-H_{x,\min})]
\end{equation} 
where 
$H_{x,\min}$ is short for $H_x(x_{\min})$, etc. and
$R$ is a random number, $R\in]0,1[$. 
The probability for keeping the point is then given by the `$g$-function',
\begin{eqnarray}
g(x,y,p_{\ell\perp},y_{\ell}) & = &
\frac{(H_{x,\max}-H_{x,\min})(H_{y,\max}-H_{y,\min})
(H_{p_{\ell\perp},\max}-H_{p_{\ell\perp},\min})
(H_{y{\ell},\max}-H_{y{\ell},\min})}
{h_{x}(x)h_{y}(y)h_{p_{\ell\perp}}(p_{\ell\perp})h_{y_{\ell}}(y_{\ell})}
\nonumber \\ & &
\times\frac{d\sigma}{dx\,dy\,dp_{\ell\perp}\,dy_{\ell}}\,
\label{g-func}
\end{eqnarray}
divided by its maximum. 
The weighting functions $h$ should be chosen
to mimic the characteristic behaviour of the differential cross-section. 
This makes the function $g$ smooth which
is important to make the rejection technique efficient.
For details on the Monte Carlo methods used, see \cite{XJOB}.

From the phase
space point ($x, y, p_{\ell\perp}, y_{\ell}$), the four-momenta
of the particles (partons) in the Feynman diagram, Fig.~1, are calculated.
The Lund Monte Carlo programs {\sc Lepto} 6.5 \cite{LEPTO} and
{\sc Jetset} 7.4 \cite{JETSET} are then used
to produce a complete final state of observable particles. 
The on-shell $W$/$Z$ (from the heavy neutrino decay)
is decayed with the proper
branching ratios into a lepton pair or a $q\bar{q}'$ pair, where the
polarisation of the $W$/$Z$ is not taken into account.
(In the case of a right-handed $W$, it is assumed to decay only hadronically
since the right-handed neutrino is assumed to be heavier than the right-handed
$W$.) For $W$/$Z$$\rightarrow q\bar{q}'$,
parton showers are included to account for QCD radiation of additional
partons, and this parton system is then hadronised using the Lund
string model \cite{LUND,JETSET}.
Similarly, the quark coming into and leaving
the deep inelastic scattering may radiate partons through initial and final
state parton showers. Together with the proton remnant spectator,
this parton system is hadronised with the Lund model. Thus, the complete
`history' of the event is generated resulting in a complete final state.
For a more complete description of the implementation, see \cite{XJOB}.

\section{Description of program components}
The program is written in FORTRAN 77 and consists of a set of subroutines
that must be activated by the users main steering program, which
should call the subroutine MAINIT to initialize the
generator and then call the subroutine MAJOR for each new event to be
generated. All subroutine and common-block names start with MA
to indicate origin and avoid name clashes.
The only exception is the real function AMGFUN which starts
with AM to follow the FORTRAN name convention.

\subsection{Subroutines and functions}

\noindent
The following subroutines should be called by the user:

\newpage
\noindent
SUBROUTINE {\bf MAINIT}
\begin{defl}{12345678901234}
\item[{\it Purpose:}]
Initiate constants and starting values. Calculate the maxima of the
$g$-function, Eq.~(\ref{g-func}), if not given by the user. {\bf Has 
to be called once before MAJOR is called.} 
\item[{\it Called by:}] 
User
\item[{\it Calls to:}]
LINIT, MADEFM, MADMAX
\end{defl}

\noindent
SUBROUTINE {\bf MAJOR}
\begin{defl}{12345678901234}
\item[{\it Purpose:}]
Administer the generation of one event, calculate the four momenta and fill 
the event record. {\bf Has to be called once for every event to be generated.}
\item[{\it Called by:}] 
User
\item[{\it Calls to:}] 
MAGENE, MAFLAV, MAERRM, LSHOWR, LUSHOW, LUPREP, LUEDIT, LUDBRB, LUROBO, LUEXEC
\item[{\it Functions used:}]
ULMASS, RLU, ULANGL, LUCOMP, PLU 
\end{defl}

\noindent
The following subroutines and functions are called internally:

\noindent
SUBROUTINE {\bf MADEFM}
\begin{defl}{12345678901234}
\item[{\it Purpose:}]
Define the Majorana neutrino in {\sc Jetset} 7.4 code (KF=79).
\item[{\it Called by:}] 
 MAINIT 
\item[{\it Calls to:}] 
-
\end{defl}

\noindent
SUBROUTINE {\bf MADMAX}
\begin{defl}{12345678901234}
\item[{\it Purpose:}]
Set appropriate starting values for MINUIT \cite{MINUIT} and call MINUIT 
for the calculation of the maxima of the $g$-function.
\item[{\it Called by:}] 
MAINIT
\item[{\it Calls to:}] 
MAINEW
\end{defl}

\noindent
SUBROUTINE {\bf MASIGX}(NPAR,DERIV,DIFSIG,XF,NFLAG)
\begin{defl}{12345678901234}
\item[{\it Purpose:}]
Is called by MINUIT and is used as an interface to the $g$-function AMGFUN.
Gives minus the $g$-function if within range and otherwise it gives the 
square  of the distance to the limit. The search for the maximum is done 
along the singularity of the $g$-function (only $x$ and $y$ varies freely
while $p_{\ell\perp}$ and $y_{\ell}$ are calculated from $x$ and $y$).
Stores the maxima found by MINUIT in the array GMAX. For details, 
see \cite{XJOB}.
\item[{\it Called by:}] 
MINUIT-subroutines 
\item[{\it Calls to:}] 
MAERRM
\item[{\it Functions used:}]
AMGFUN
\end{defl}

\noindent
SUBROUTINE {\bf MAGENE}
\begin{defl}{12345678901234}
\item[{\it Purpose:}]
Generate a phase space point according to the differential cross-section.
\item[{\it Called by:}] 
MAJOR
\item[{\it Calls to:}] 
 -
\item[{\it Functions used:}]
RLU, AMGFUN
\end{defl}

\noindent
REAL FUNCTION {\bf AMGFUN}(C)
\begin{defl}{12345678901234}
\item[{\it Purpose:}]
Calculate the value of the $g$-function, Eq.~(\ref{g-func}), in a given phase 
space point.
\item[{\it Called by:}] 
MASIGX, MAGENE
\item[{\it Calls to:}] 
PYSTFU
\end{defl}

\noindent
SUBROUTINE {\bf MAFLAV}(NINITQ,NSCATQ)
\begin{defl}{12345678901234}
\item[{\it Purpose:}]
Choose flavour of initial quark and scattered quark at the boson
vertex according to the relative cross-sections and the CKM matrix.
\item[{\it Called by:}] 
MAJOR
\item[{\it Calls to:}] 
 -
\item[{\it Functions used:}]
RLU, LUCOMP
\end{defl}

\noindent
SUBROUTINE {\bf MAERRM}(CHERRM,ERRVAL)
\begin{defl}{12345678901234}
\item[{\it Purpose:}]
Handle errors and write error messages.
\item[{\it Called by:}] 
MAJOR, MASIGX, MAGENE, AMGFUN
\item[{\it Calls to:}] 
 - 
\end{defl}

\noindent
The following subroutines are modified versions of MINUIT subroutines
from LEPTO: MACMND, MAIDAT, MAINEW, MAINTO, MAPRIN, MARAZZ and MASIMP.
In addition the following subroutines (S) and functions (F) in 
{\sc Lepto} 6.5 (L) and {\sc Jetset} 7.4 (J) are used:
\begin{defl}{123456789012345678}
\item[{\it Routine}]{\it Purpose}
\item[LINIT (S/L)]   Initialize {\sc Lepto}
\item[LSHOWR (S/L)]  Include parton cascades
\item[LYSTFU (S/L)]  Give the parton distribution functions
\item[LUDBRB (S/J)]  Perform rotation and boost in double precision
\item[LUROBO (S/J)]  Perform rotation and boost in single precision
\item[LUSHOW (S/J)]  Generate timelike parton showers
\item[LUPREP (S/J)]  Rearrange parton shower end products
\item[LUEDIT (S/J)]  Exclude unstable or undetectable jets/particles 
from the event record
\item[LUEXEC (S/J)]  Administrate the fragmentation and decay chain
\item[ULMASS (F/J)]  Give the mass of a parton/particle
\item[RLU (F/J)]     Generate a (pseudo)random number uniformly in 
$0<{\rm RLU}<1$
\item[ULANGL (F/J)]  Calculate the angle from the $x$ and $y$ coordinates
\item[LUCOMP (F/J)]  Give the compressed parton/particle code
\item[PLU (F/J)]     Provide various real-valued event data
\end{defl}

\subsection{Common blocks}

The common-block mainly intended for communication with the program is
MAUSER, which sets switches, parameters and cuts. In addition,
all kinematic variables for a given event can be found in MAKINE and
LUJETS in {\sc Jetset} 7.4 is used to store the event record.
All variables are given sensible default values in the
block data MADATA and all variable names obey the following name
convention: integers start with I-N, single precision reals start with A-C,E-H
and O-Z and double precision reals start with D.

\noindent
COMMON /{\bf MAUSER}/ MAFLAG(20),CUTM(12),PARM(30)
\begin{defl}{1234567890123456}
\item[{}]
{\it Switches for controlling the program:} 
\item[MAFLAG(1)]
(D=9) choice of parameterisation for the parton densities in the proton.
Is transferred to LST(15) in {\sc Lepto} \cite{LEPTO} which in version 
6.5 gives the following choices of relevance. (For information on how to use
PDFLIB \cite{pdflib}, see  LST(16) in {\sc Lepto}.The parton densities 
are obtained from subroutine PYSTFU in {\sc Pythia} 5.7 \cite{JETSET}.)

\item[{\hfill =0:}] 
parton density choice and parameters are controlled
     directly through parameters in {\sc Pythia 5.7}.
\item[{\hfill =1:}]  Eichten-Hinchliffe-Lane-Quigg set 1 \cite{EHLQ}.
\item[{\hfill =2:}]  Eichten-Hinchliffe-Lane-Quigg set 2 \cite{EHLQ}.
\item[{\hfill =3:}]  Duke-Owens set 1 \cite{DO}.
\item[{\hfill =4:}]  Duke-Owens set 2 \cite{DO}.
\item[{\hfill =5:}]  CTEQ2M (best $\overline{MS}$ fit) \cite{CTEQ}.
\item[{\hfill =6:}]  CTEQ2MS (singular at small-$x$) \cite{CTEQ}.
\item[{\hfill =7:}]  CTEQ2MF (flat at small-$x$) \cite{CTEQ}.
\item[{\hfill =8:}]  CTEQ2ML (large $\Lambda$)) \cite{CTEQ}.
\item[{\hfill =9:}]  CTEQ2L (best leading order fit) \cite{CTEQ}.
\item[{\hfill =10:}]  CTEQ2D (best DIS fit) \cite{CTEQ}.
\item[MAFLAG(2)]
(D=0) choice of factorisation scale $\mu^2$ used in the structure-functions
\item[{\hfill =0:}]
$Q^2$
\item[{\hfill =1:}]
$p_{N\perp}^2$, the squared transverse momentum of the neutrino
\item[{\hfill =2:}] 
the constant given in PARM(4)
\item[MAFLAG(3)]
(D=0) regulates the final-state lepton status in the event record, K(I,1)
\item[{\hfill =0:}]
active final-state lepton, i.e.~K(I,1)=1
\item[{\hfill =1:}]
inactive final-state lepton, i.e.~K(I,1)=21
\item[MAFLAG(4)]
(D=1) regulates the direction of the $x$-axis ($z$-axis in proton direction)
\item[{\hfill =0:}]
the $x$-axis along $p_{N\perp}$
\item[{\hfill =1:}]
event rotated randomly in azimuthal angle between 0 and 2$\pi$
\item[MAFLAG(5)]
(D=1) regulates the  hadronisation
\item[{\hfill =0:}]
hadronisation off
\item[{\hfill =1:}]
hadronisation on
\item[MAFLAG(6)]
(D=0) regulates the amount of written output.
\item[{\hfill =0:}]
all output except MINUIT-output 
\item[{\hfill =5:}]
all output
\item[MAFLAG(7)]
(D=0) regulates the choice of the process number (NPROC)
\item[{\hfill =0:}]
1, 2 and 3 mixed according to their respective branching ratios
\item[{\hfill =1:}]
$e^- p \rightarrow {N} X \rightarrow{e}^+ W^- X$
\item[{\hfill =2:}]
$e^- p \rightarrow {N} X \rightarrow{e}^- W^+ X$
\item[{\hfill =3:}]
$e^- p \rightarrow {N} X \rightarrow \nu Z X$
\item[{\hfill =10:}]
11, 12 and 13 mixed according to their respective branching ratios
\item[{\hfill =11:}]
$e^+ p \rightarrow {N} X \rightarrow{e}^- W^+ X$ 
\item[{\hfill =12:}]
$e^+ p \rightarrow {N} X \rightarrow{e}^+ W^- X$
\item[{\hfill =13:}]
$e^+ p \rightarrow {N} X \rightarrow \nu Z X$
\item[MAFLAG(8)]
(D=2) simulation of QCD effects in the scattered quark and proton remnant 
system (for 2-5 {\it cf.} LST(8) in {\sc Lepto} \cite{LEPTO})
\item[{\hfill =1:}]
final state radiation from the scattered quark only, using a simplified 
treatment (LUSHOW is used on the quark and diquark system). Only valence 
quarks, {\it i.e.}u-quarks, are considered in the proton.
\item[{\hfill =2:}]
initial and final state radiation
\item[{\hfill =3:}]
initial state radiation
\item[{\hfill =4:}]
final state radiation
\item[{\hfill =5:}]
QCD switched off but target remnant as 2, 3 and 4

\item[MAFLAG(9)]
error flag
\item[{\hfill =1:}]
maximum of $g$-function violated. This is {\bf serious} if the violation is 
large and in that case the user has to either set the maxima by hand 
(see MAFLAG(19)) or change the $h$-functions to make the $g$-function 
smoother
\item[MAFLAG(10)]
(D=1) Error handling  
\item[{\hfill =0:}]
no warnings and execution not stopped on error 
\item[{\hfill =1:}]
warnings printed but execution not stopped 
\item[{\hfill =2:}]
warnings printed and execution stopped 
\item[MAFLAG(11)] 
(D=2) choice of $h_x$-function 
(see Section 2 for definition and Table \ref{tab1} for suitable choices) 
\item[{\hfill =-2:}]
$h_x(x)=1/x^2$ 
\item[{\hfill =-1:}]
$h_x(x)=1/x$ 
\item[{\hfill =0:}]
$h_x(x)=const.$ 
\item[{\hfill =1:}]
$h_x(x)=\exp(-A_xx)$ 
\item[{\hfill =2:}]
$h_x(x)=x\exp(-A_xx^2)$ 
\item[MAFLAG(12)] 
(D=0) regulates calculation of $A_x$ 
\item[{\hfill =0:}]
$A_x$ is calculated by the program  
\item[{\hfill =1:}]
$A_x$ is given in PARM(8) 
\item[MAFLAG(13)] 
(D=1) choice of $h_y$-function 
(see Section 2 for definition and Table \ref{tab1} for suitable choices) 
\item[{\hfill =-2:}]
$h_y(y)=1/y^2$ 
\item[{\hfill =-1:}]
$h_y(y)=1/y$ 
\item[{\hfill =0:}]
$h_y(y)=const.$ 
\item[{\hfill =1:}]
$h_y(y)=\exp(-A_yy)$ 
\item[{\hfill =2:}]
$h_y(y)=y\exp(-A_yy^2)$ 
\item[MAFLAG(14)] 
(D=0) regulates calculation of $A_y$ 
\item[{\hfill =0:}]
$A_y$ is calculated by the program  
\item[{\hfill =1:}]
$A_y$ is given in PARM(9) 
\item[MAFLAG(15)] 
(D=2) choice of $h_{p_{\ell\perp}}$-function 
(see Section 2 for definition and Table \ref{tab1} for suitable choices) 
\item[{\hfill =-2:}]
$h_{p_{\ell\perp}}({p_{\ell\perp}})=1/{p_{\ell\perp}}^2$ 
\item[{\hfill =-1:}]
$h_{p_{\ell\perp}}({p_{\ell\perp}})=1/{p_{\ell\perp}}$ 
\item[{\hfill =0:}]
$h_{p_{\ell\perp}}({p_{\ell\perp}})=const.$ 
\item[{\hfill =1:}]
$h_{p_{\ell\perp}}({p_{\ell\perp}})=
\exp(-A_{p_{\ell\perp}}{p_{\ell\perp}})$ 
\item[{\hfill =2:}]
gives $h_{p_{\ell\perp}}({p_{\ell\perp}})=
{p_{\ell\perp}}\exp(-A_{p_{\ell\perp}}{p_{\ell\perp}}^2)$ 
\item[MAFLAG(16)] 
(D=0) regulates calculation of $A_{p_{\ell\perp}}$
 for ${N}\rightarrow{e}^{\pm} W^{\mp}$  
\item[{\hfill =0:}]
$A_{p_{\ell\perp}}$ is calculated by the program  
\item[{\hfill =1:}]
$A_{p_{\ell\perp}}$ is given in PARM(10) 
\item[MAFLAG(17)] 
(D=0) regulates calculation of $A_{p_{\ell\perp}}$
for ${N}\rightarrow{\nu} Z$  
\item[{\hfill =0:}]
$A_{p_{\ell\perp}}$ is calculated by the program  
\item[{\hfill =1:}]
$A_{p_{\ell\perp}}$ is given in PARM(11) 
\item[MAFLAG(18)] 
(D=1) choice between fixed and varying (according to a 
            Breit-Wigner
  distribution) boson-masses in the neutrino decay 
\item[{\hfill =0:}]
fixed boson-masses 
\item[{\hfill =1:}]
varying boson-masses 
\item[MAFLAG(19)] 
(D=0) regulates calculation of $g$-function 
(see section 2) maxima  
\item[{\hfill =0:}]
maxima of $g$-functions are calculated 
\item[{\hfill =1:}]
maxima of $g$-functions given in PARM(12) to PARM(17) 
\item[MAFLAG(20)] 
(D=0) choice between left and right handed $W$ exchange 
\item[{\hfill =0:}]
left handed $W$ exchange 
\item[{\hfill =1:}]
right handed $W$ exchange 
\end{defl}

\begin{defl}{12345678901}
\item[{}]
{\it Cuts on kinematic variables defined in Section 1:}  
\item[CUTM(1):]
$x_{\min}$ (D=0.)  
\item[CUTM(2):] 
$x_{\max}$ (D=1.)  
\item[CUTM(3):] 
$y_{\min}$ (D=0.)  
\item[CUTM(4):] 
$y_{\max}$ (D=1.)  
\item[CUTM(5):] 
$Q^2_{\min}$ (D=4. ${\rm GeV}^2$)  
\item[CUTM(6):] 
$Q^2_{\max}$ (D=$10^8$ ${\rm GeV}^2$)  
\item[CUTM(7):] 
$W^2_{\min}$ (D=9. ${\rm GeV}^2$)  
\item[CUTM(8):] 
$W^2_{\max}$ (D=$10^8$ ${\rm GeV}^2$)  
\item[CUTM(9):] 
$p_{\ell\perp\min}$ (D=0. GeV)  
\item[CUTM(10):] 
$p_{\ell\perp\max}$ (D=1000 GeV)  
\item[CUTM(11):] 
$y_{\ell\min}$ (D=$-100$)  
\item[CUTM(12):] 
$y_{\ell\max}$ (D=100)  
\end{defl}
 
\begin{defl}{12345678901}
\item[{}]
{\it Parameters for input (1-20) and output (21-30):}  
\item[PARM(1):]
(D=100 GeV) mass of heavy Majorana neutrino
\item[PARM(2):] 
(D=820 GeV) proton-momentum
\item[PARM(3):] 
(D=30 GeV) electron-momentum
\item[PARM(4):] 
(D=1000 ${\rm GeV}^2$)  fixed factorisation scale $\mu^2$ 
 used if MAFLAG(2)=2  
\item[PARM(5):] 
(D=0.01) degree of mixing between light and heavy Majorana
neutrinos, $|(V\xi)_{eN}|^2=|\xi_{\nu_e N}|^2$.     
In the case of right handed $W$ exchange it is interpreted as the ratio
between the right and left handed gauge couplings, $(g_R/g_L)^2$ 
which normally is assumed to be 1. 
\item[PARM(6):] 
(D=0.001) cut used around the singularity of the 
$g$-function which divides the $g$-function  into two parts 
(see ALSCUT in \cite{XJOB} for details). A larger cut will
speed up    the simulation but at the same time increase the risk of 
violating the maximum  of the $g$-function and cut away 
a larger part of the phase-space.
\item[PARM(7):] 
(D=1.1) safety factor, multiplies the maxima of the $g$-function.
\item[PARM(8):]  
$A_x$ (see also MAFLAG(12)) 
\item[PARM(9):]  
$A_y$ (see also MAFLAG(14)) 
\item[PARM(10):]  
$A_{p_{\ell\perp}}$ for ${N}\rightarrow{e}^{\pm} W^{\mp}$ 
(see also MAFLAG(16)) 
\item[PARM(11):]  
$A_{p_{\ell\perp}}$ for ${N}\rightarrow{\nu} Z$ 
(see also MAFLAG(17)) 
\item[PARM(12):] 
Max of first part of $g$-function for
 process nr 1 or 11 (see also MAFLAG(19)) 
\item[PARM(13):] 
Max of second part of $g$-function for 
 process nr 1 or 11 (see also MAFLAG(19)) 
\item[PARM(14):] 
Max of first part of $g$-function for 
 process nr 2 or 12 (see also MAFLAG(19)) 
\item[PARM(15):] 
Max of second part of $g$-function for 
 process nr 2 or 12 (see also MAFLAG(19)) 
\item[PARM(16):] 
Max of first part of $g$-function for 
 process nr 3 or 13 (see also MAFLAG(19)) 
\item[PARM(17):] 
Max of second part of $g$-function for 
 process nr 3 or 13 (see also MAFLAG(19)) 
\item[PARM(18):]  
Mass of right-handed $W$-boson in GeV  
\item[PARM(19):]  
Width of right-handed $W$-boson in GeV  
\item[PARM(20):]  
not used  
\item[PARM(21):] 
Estimate\footnotemark[1] 
of the cross-section in $pb$ 
(including cuts) for process nr 1 or 11. 
\item[PARM(22):] 
Estimate\footnotemark[1]
 of the standard deviation in PARM(21) 
\item[PARM(23):] 
Estimate\footnotemark[1] of the cross-section in $pb$ 
(including cuts) for process nr 2 or 12.  
\item[PARM(24):] 
Estimate\footnotemark[1] of the standard deviation in PARM(23) 
\item[PARM(25):] 
Estimate\footnotemark[1] of the cross-section in $pb$ 
(including cuts) for process nr 3 or 13.  
\item[PARM(26):] 
Estimate\footnotemark[1] of the standard deviation in PARM(25) 
\item[PARM(27):] 
Efficiency\footnotemark[2] in the generation 
for process nr 1 or 11   
\item[PARM(28):] 
Efficiency\footnotemark[2] in the generation 
for process nr 2 or 12   
\item[PARM(29):] 
Efficiency\footnotemark[2] in the generation 
for process nr 3 or 13   
\item[PARM(30):] 
 not used 
\end{defl}
\footnotetext[1]{The estimate is
  updated for each event so it
  should not be used until all events have been generated.}
\footnotetext[2]{The efficiency is
  given by the number of accepted events divided by the number of tries.}

\noindent
COMMON /{\bf MAKINE}/ EP,EE,S,SHAT,Q2,W2,X,Y,PNT,PNL,YN,EN,PLT,YL
\begin{defl}{12345678}
\item[{}]
{\it Kinematical variables in a given event of use for the user:}  
\item[EP:]  
Proton energy in GeV  
\item[EE:]   
Electron energy in GeV 
\item[S:]  
 CMS-energy squared (Mandelstam $s$) in GeV$^2$ 
\item[SHAT:]   
$\hat{s}=x\,{s}$ 
\item[Q2:]   
Momentum transfer squared, $Q^2=-q^2$ in GeV$^2$ 
\item[W2:]   
Hadronic CMS-energy squared, $W^2=(p_p+q)^2$ in GeV$^2$ 
\item[X:]   
Bjorken-$x$, $x=Q^2/2p_p\cdot{q}$ 
\item[Y:]   
Standard $y$-variable, $y=p_p\cdot{q}/p_p\cdot{p_e}$ 
\item[PNT:]   
Transverse momentum of heavy $p_{N\perp}$ Majorana neutrino in GeV  
\item[PNL:]   
Longitudinal momentum of heavy Majorana neutrino in GeV  
\item[YN:]   
Rapidity $y_N$ of heavy Majorana neutrino 
\item[EN:]   
Energy of heavy Majorana neutrino in GeV  
\item[PLT:]   
Transverse momentum $p_{\ell\perp}$ of final-state lepton 
from $N$ decay in GeV  
\item[YL:]   
Rapidity $y_{\ell}$ of final-state lepton 
\end{defl}
 
\noindent
COMMON /{\bf LUJETS}/ N,K(4000,5),P(4000,5),V(4000,5) \\
\noindent
The variables in the common-block LUJETS are described in the 
{\sc Jetset} 7.4 manual \cite{JETSET}. The first seven entries in the 
event record are as follows
(in the lab frame with the $z$-axis in the proton direction):
\begin{defl}{1234567890}
\item[\hfill  1.]  Incoming electron  
\item[\hfill  2.]  Incoming proton  
\item[\hfill  3.]  Exchanged $W$-boson  
\item[\hfill  4.]  Heavy Majorana neutrino  
\item[\hfill  5.]  Incoming parton before initial shower  
\item[\hfill  6.]  Incoming quark at boson vertex  
\item[\hfill  7.]  Scattered quark at boson vertex before final shower  
\end{defl}
\noindent
The decay products from the $N$-decay are in line K(4,4) (final state lepton)
and   K(4,5) (on shell boson) respectively.
 
\noindent
Common-blocks for internal use:
MACROS, MAMAMI, MAMASS, MACONS, MAGSPE.\\
\noindent
{\sc Lepto} 6.5 common-blocks used :
LEPTOU, LBOOST, LFLMIX, LPFLAG, LYPARA.\\
\noindent
{\sc Jetset} 7.4  common-blocks used :
LUDAT1, LUDAT2, LUDAT3, LUDAT4.

\subsection{Update history}
Updates 
from version 1.1 \cite{major11} to 1.3 \cite{major13}:
\begin{itemize}
 \item inclusion of the neutral current decay $N\rightarrow\nu_{\ell}Z$
 \item extension and redefinition of the common-block MAUSER and 
       some internal common-blocks
 \item inclusion of a subroutine for error handling
 \item correction of a small error in the differential cross-section
\end{itemize}
from version 1.3 \cite{major13} to 1.5:
\begin{itemize}
 \item inclusion of possibility to have incoming positron (see MAFLAG(7))
 \item inclusion of possibility to have exchange of right-handed W's
 which are lighter than the Heavy Neutrino (see MAFLAG(20), PARM(5), 
 PARM(18) and PARM(19))
 \item update to run with {\sc Lepto} 6.5, {\sc Jetset} 7.4 and 
 {\sc Pythia} 5.7 (see MAFLAG(1))
 \item inclusion of MINUIT \cite{MINUIT} subroutines from {\sc Lepto}
\end{itemize}

\section{Usage and availability}
{\sc Major} 1.5 should be loaded together with {\sc Lepto} 6.5 
\cite{LEPTO}, {\sc Jetset} 7.4 and {\sc Pythia} 5.7 \cite{JETSET}. 
The program is a slave system, which the user must call from
his own steering program.
Information about the program, the source code and a demonstration job 
can be found via the WWW on the {\sc Major} homepage 
\texttt{ http://www3.tsl.uu.se/thep/major/} 
or from one of the authors via email.
It is {\bf not} recommended to split the program 
and making a library of it since this 
can give problems with the use of {\sc Lepto} routines.
Examples of suitable choices of $h$-functions are given
in Table \ref{tab1}.

\begin{table}
\caption{Examples of suitable choices of $h$-functions at different 
{\it cms} energies and for different heavy Majorana neutrino masses.
\label{tab1}}
\begin{tabular}[h]{|l|c|cc|}
\hline
\hline
                & HERA & \multicolumn{2}{c|}{LEP$\otimes$LHC}  \\
$\sqrt{s}$ [GeV]& 300  & \multicolumn{2}{c|}{1200}  \\
$m_N$ [GeV]     & 100 & 300 & 700  \\
\hline
MAFLAG(11)      & 2   & -1  & -2   \\
MAFLAG(13)      & 1   & -1  & -1   \\
MAFLAG(15)      & 2   &  2  &  2   \\
\hline
\hline
\end{tabular}
\end{table}

%

\end{document}